\documentclass[pdftex,twocolumn,epjc3]{svjour3}          

\RequirePackage[T1]{fontenc}

\smartqed  

\RequirePackage{graphicx}
\RequirePackage{mathptmx}      
\RequirePackage{flushend}
\RequirePackage[numbers,sort&compress]{natbib}
\RequirePackage[colorlinks,citecolor=blue,urlcolor=blue,linkcolor=blue]{hyperref}

\journalname{Eur. Phys. J. C}

\begin{document}

\title{Holographic Dark Energy Models and Higher Order Generalizations 
in Dynamical Chern-Simons Modified Gravity
}


\author{Antonio Pasqua\thanksref{e1,addr1}
        \and
        Rold\~ao da Rocha\thanksref{e2,addr2,addr3} 
        \and Surajit Chattopadhyay
\thanksref{e3,addr4}         
}

\thankstext{e1}{e-mail: Antonio.Pasqua@univ.trieste.it,toto.pasqua@gmail.com}
\thankstext{e2}{e-mail: roldao.rocha@ufabc.edu.br}
\thankstext{e3}{e-mail:surajcha@iucaa.ernet.in}
\institute{Department of Physics,
University of Trieste, Trieste, Italy.
\label{addr1}
\and 
Centro de Matem\'atica, Computa\c c\~ao e Cogni\c c\~ao,
Universidade Federal do ABC, 09210-170, Santo Andr\'e, SP, Brazil\label{addr2}
                   \and
          International School for Advanced Studies (SISSA), Via Bonomea 265, 34136 Trieste, Italy\label{addr3}
          \and
          Pailan College of Management and Technology, Bengal
Pailan Park, Kolkata-700 104, India.\label{addr4}
}

\date{Received: date / Accepted: date}

\maketitle

\begin{abstract}
Dark Energy models are here investigated and stu\-died in the framework of the Chern-Simons modified gravity model. We bring into focus  the Holographic Dark Energy (HDE)  model with Granda-Oliveros cut-off, the Modified Holographic Ricci Dark Energy (MHRDE) model and, moreover, a model with higher derivatives of the Hubble parameter as well. The relevant expressions of the scale factor $a(t)$ for a Friedmann-Robertson-Walker  Universe  are derived and studied, and in this context, the evolution of the scale factor is shown to be similar to that one displayed by the modified Chaplygin gas in two of the above models.
\end{abstract}

\section{Introduction}
Cosmological data obtained from different independent observations of SNeIa, CMB radiation anisotropies, X-ray experiments and Large Scale Structures are well known to point toward the accelerated phase of expansion of the Universe \cite{1-1,1a,1b,cmb1,planck,tegmark,xray}. 

The cosmological constant $\Lambda$ model, Dark Energy (DE) models and theories of modified gravity, among other attempts, have been approached in order to  provide an explanation for the  accelerated expansion of the Universe   \cite{Peebles,sahni1,copeland-2006,clift,bambaDE}. The cosmological constant $\Lambda$ stands for  the most straightforward candidate suggested to explain the observational evidence for it. The fine-tuning and the cosmic coincidence problems are  questions still underlying the cosmological constant model  \cite{delcampo,delcampoh}. 

A model for DE, motivated by the holographic principle, was proposed \cite{Horava}  and further studied in \cite{3a,3b,4,5a,rami1,rami3,cardenas,wu2008,zim,0505}. 
The holographic model of DE (HDE) has been in addition 
comprehensively investigated  
\cite{gong2,noji1, noji2, noji3, noji4, noji5, noji6,li997,li998}. 
  The HDE model has been further  employed to drive inflation of the Universe  \cite{10}, and considered in \cite{12a,12e,noji4,13a,13c}  with different IR cut-offs. For example the  future event and the particle horizons and the Hubble horizon were considered as well. Moreover, correspondences between the HDE model and other scalar field models have been recently suggested \cite{14}. The HDE model fits the cosmological data by CMB radiation anisotropies and SNeIa \cite{16,16d}.

Recently, the cosmic acceleration has been also well studied by a promising modified gravity model that has recently attained visibility: the modified gravity Chern-Simons model  \cite{3cs}. The low-energy limit of string theory comprises a correction that cancels the anomalies  to the Einstein-Hilbert action, wherein  the Chern-Simons modified gravity is derived as an effective theory. Gravitational parity 
violation was first investigated using this framework   \cite{3cs}, appearing both in 4D compactifications of perturbative string 
theory and further in 
loop quantum gravity as well, when the Barbero-Immirzi parameter emulates a scalar field coupled to the Nieh-Yan 
invariant~\cite{taveras,calcagni,mercuri,rocha}. In the Chern-Simons modified gravity, the  Pontryagin topological invariant is well known  not to affect the field
equations. Thus the so called Chern-Simons correction consists of the product of the Pontryagin density by a scalar field, regarded  as either a dynamical evolving field or a non-dynamical  background field. In the former case, the dynamical Chern-Simons modified gravity (DCSMG) is therefore approached \cite{6cs,13cs}.  
Some efforts have recently  provided  bounds to the Chern-Simons parameter, accordingly  \cite{csca}.

In this paper, we study the many faces of DE models in the context of the dynamic formulation of Chern-Simons gravity, where the coupling constant is promoted to a scalar field. Recent applications include for instance neutron star binary \cite{7cs}. We shall investigate three different DE models: the HDE model with Granda-Oliveros (GO) cut-off, the Modified Holographic  Ricci Dark Energy (MHRDE) model and a recently proposed model with higher derivatives of the Hubble parameter $H$ \cite{34cs} in the framework of Chern-Simons gravity, in order to obtain the expressions of the scale factor for each model. We prove that  both 
the HDE model with GO cut-off and the model with higher derivatives of the Hubble parameter $H$  in the framework gravity Chern-Simons are related to the modified Chaplygin gas models \cite{cgas3,cgas41,cgas42} that further represent  the well known models of dark energy as Chaplygin gas \cite{cga,cgas0,cgask}.

The paper is organized as follows. In Section II, we briefly revisit dynamical Chern-Simons modified gravity model. In Section III, we describe the three different DE models considered in this work in the framework of the Chern-Simons modified gravity models and we derive the relevant expressions of the scale factor $a\left( t \right)$. Finally, in Section IV, we write the conclusions of this work.

\section{Chern-Simons Gravity}
This section is devoted to provide the main features of the Chern-Simons modified gravity model. A homogeneous and isotropic Universe described by the Friedmann-Robertson-Walker (FRW) metric  is governed by the metric 
\begin{eqnarray}
ds^2 = -dt^2 + a^2\left( t \right) \left( \frac{dr^2}{1-kr^2} + r^2d\theta^2 +r^2 \sin ^2 \theta d\phi^2 \right), \label{metricfrw}
\end{eqnarray}
where $a\left(t\right)$ is the scale factor and $k$ denotes the curvature parameter assuming the values +1, 0 and $-1$  leading respectively to an open, a flat or a closed Universe.  
The action for the Chern-Simons (CS)  theory is given by \cite{6cs,csca,7cs,13cs}:
\begin{eqnarray}
S &=& \frac{1}{16\pi G} \int d^4 x \left[ \sqrt{-g}R + \frac{\ell}{4}\theta \,^{\star}\! R^{\rho\sigma\mu\nu} R_{\sigma\rho\mu\nu}\right.\nonumber\\&& \left. \qquad\qquad- \frac{1}{2} g^{\mu\nu}  \nabla_{\mu} \theta \nabla_\nu\theta
 + V(\theta)   \right] + S_{mat},\label{action}
\end{eqnarray}
where $G$ represents the Newton's gravitational constant, $g$ is the determinant of the spacetime metric $g_{\mu\nu}$,  $R= g^{\mu\nu} R_{\mu\nu} $ denotes the 
Ricci scalar defined with the Ricci tensor
$R_{\mu\nu}=R^{\rho}_{\ \mu\rho\nu} $, and 
$R^{\mu}_{\ \nu\rho\sigma}$
stands for the  Riemann tensor components and
$ ^{\star}\! R^{\rho\sigma\mu\nu} := \frac{1}{2} 
 \varepsilon^{\mu\nu\alpha\beta} R^{\rho\sigma}_{\quad \alpha\beta}$
denotes the components of the Hodge dual Riemann tensor. The term $\ell$ denotes the 4D coupling constant,  $\,^{\star}\! R^{\rho\sigma\mu\nu} R_{\sigma\tau\mu\nu}$  is the Pontryagin invariant, and 
the function $\theta$ indicates the dynamical scalar field of the model and $S_{mat}$ represents the action of matter.   For simplicity we consider the potential $V(\theta)$ equal to zero. 
The second term, which is called the Chern-Pontryagin 
(CP) term, can be converted to the CS term 
via partial integration as 
\[
 -\frac{\ell}{32\pi G} \int d^4 x \sqrt{-g} 
 \varepsilon^{\mu\nu\sigma\rho} \partial_{\mu} \theta 
\left( \Gamma^{\alpha}_{\nu \beta} \partial_{\sigma} 
  \Gamma^{\beta}_{\rho\alpha} + \frac{2}{3} 
  \Gamma^{\tau}_{\nu\alpha} 
  \Gamma^{\alpha}_{\sigma\beta} \Gamma^{\beta}_{\tau\rho} 
  \right) .\nonumber
\]
The third term in Eq.~(\ref{action}) is the kinematic term 
for $\theta$. 
By varying the action $S$ given in Eq. (\ref{action}) with respect to the metric tensor $g_{\mu \nu}$ and to the scalar field $\theta $,  the following field equations are respectively obtained:
\begin{eqnarray}
G_{\mu \nu} + \ell C_{\mu \nu} &=& 8\pi G T_{\mu \nu}, \label{field1}\\
g^{\mu \nu}\nabla_{\mu} \nabla_{\nu} \theta &=& -\frac{\ell}{64\pi}\,  ^{\star} R^{\rho\sigma\mu\nu} R_{\sigma\tau\mu\nu},\label{field2}
\end{eqnarray}
where $G_{\mu \nu}$ represents the Einstein tensor components and $C_{\mu \nu}$ indicates the Cotton tensor:
\begin{eqnarray}
 C^{\mu\nu} 
 &=& -\frac{1}{2\sqrt{\!-g}}\left(\left( \nabla_{\rho} \theta \right) 
  \!   \varepsilon^{\rho\beta\tau (\mu} \nabla_{\tau} 
     R^{\nu)}_{\ \beta}\right. \nonumber\\&&\! 
\qquad\qquad + \left. \left( \nabla_{\sigma} \nabla_{\rho} \theta \right)
   \!\!\: ^{\star}\! R^{\rho ( \mu\nu ) \sigma}\right)\,. 
 \label{cotton}
\end{eqnarray}
The energy-momentum tensor $T_{\mu \nu}= \mathring{T}_{\mu \nu} + T_{\mu \nu}^{\theta}$ has two terms: 
\begin{eqnarray}
T_{\mu \nu}^{\theta} &=& -\frac{1}{2}g_{\mu \nu}\nabla^{\rho} \theta    \nabla_{\rho} \theta + \nabla_{\mu} \theta      \nabla_{\nu} \theta  ,\label{emom2}
\\\mathring{T}_{\mu \nu} &=& \mathring{p}\, g_{\mu \nu}+ \left( \mathring{\rho} + \mathring{p}  \right)U_{\mu} U_{\nu} \,. \label{emom3}
\end{eqnarray}
The term $T_{\mu \nu}^{\theta}$ represents the scalar field contribution and $\mathring{T}_{\mu \nu}$ 
indicates  the energy-momentum tensor of the corresponding DE model. Furthermore,  $U_{\mu} = \left(1, 0, 0, 0\right)$
 denotes the standard time-like 4-velocity, $\mathring{\rho}$ represents the energy density   and $\mathring{p}$ stands for  the pressure of the DE model to be  considered. 
 The component $T_{00}$ provides the Friedmann equation
\begin{eqnarray}
G_{00} + C_{00} = \mathring{T}_{00} + T_{00}^{\theta}, \label{fri1}
\end{eqnarray}
for 
\begin{eqnarray}
G_{00} &=& 3 \left( \frac{\dot{a}^2}{a^2} + \frac{k}{a^2}  \right),\quad
T_{00}^{\theta} = \frac{1}{2}\dot{\theta}^2, \qquad
\mathring{T}_{00} = \mathring{\rho},
\end{eqnarray}
where $(\;\,\dot{}\;\,)=d/dt$. 
Moreover, as the components of the Cotton tensor vanish for all
spherically symmetric metrics \cite{13cs}, in particular  $C_{00} = 0$. Therefore, the Friedmann equation reads
\begin{eqnarray}
\frac{\dot{a}^2}{a^2} + \frac{k}{a^2} = \frac{1}{3}\mathring{\rho} + \frac{1}{6}\dot{\theta}^2. \label{frinew}
\end{eqnarray}
It is worth  to emphasize that Eq. (\ref{frinew}) takes into account $8\pi G=1$.

In what follows we study Eq. (\ref{frinew}) in the context of the modified Chern-Simons modified gravity for  three different DE models, namely, the HDE model with Granda-Oliveros cut-off, the MHRDE model and a recent DE model     which involves the Hubble parameter squared and the first and the second time derivatives of the Hubble parameter. We shall derive an expression of the scale factor for each one of these models.

\section{Dark Energy Models in Chern-Simons Gravity}
In this Section, our aim is to give a brief description of the DE models dealt with and to study their behavior in the framework of Chern-Simons modified gravity model, in order to find the expressions of the scale factor $a(t)$.
In the first subsection, we will consider the HDE model with Granda-Oliveros cut-off, in the second subsection the MHRDE model while in the third one the model with higher derivatives of the Hubble parameter $H$.

\subsection{The HDE Model with GO Cut-off}
Granda and Oliveros introduced a new IR cut-off, by including a term proportional to $\dot{H}$ and one term proportional to $H^2$. This new IR cut-off  $L_{GO}$ is known as Granda-Oliveros (GO) scale, provided by \cite{grandaoliveroscs,grandaoliverosacs}:
\begin{equation}
L_{GO}=\left( \alpha\dot{H}+\beta H^{2}\right) ^{-1/2},  \label{lgo5}
\end{equation}
where $\alpha $ and $\beta $ represent two  constant parameters.
As the underlying origin of the HDE model lacks, the term with the time derivative of the Hubble parameter is expected, since this term appears in the curvature scalar \cite{grandaoliverosacs,grandaoliveroscs}. 

The expression of the HDE energy density with $L_{GO}$ cut-off is given by:
\begin{equation}
\rho_{D_{GO}}= 3c^2\left(\alpha \dot{H}+ \beta H^{2}\right)\,,  \label{lgo5-1}
\end{equation} where the  numerical
constant $c$ arises from observational data. For  a flat [non-flat] 
Universe  we have $c=0.818_{-0.097}^{+0.113}$  
$\left[c=0.815_{-0.139}^{+0.179}\right]$ \cite{li997,li998}.  
It is worth  to emphasize that the Planck mass $M_p$ is considered hereon normalized to the unit. 
In the limiting case corresponding to $ \alpha = 1$ and $\beta = 2$, the scale $L_{GO}$ reduces to the average radius of the Ricci scalar curvature, when $k=0$ in Eq.(\ref{metricfrw}), corresponding to a flat Universe.
 DE models that consider the GO scale avoid  the causality problem.

The Friedmann equation  (\ref{frinew}), when the  energy density of DE in Eq. (\ref{lgo5-1}) is taken into account, reads (here we consider the normalization $\rho_{D_{GO}}/c^2\mapsto\rho_{D_{GO}}$):
\begin{eqnarray}
\frac{\dot{a}^2}{a^2} + \frac{k}{a^2} = \alpha \dot{H} + \beta H^2 + \frac{1}{6}\dot{\theta}^2\,,\end{eqnarray} which can be written by substituting $
\dot{H} = \frac{\ddot{a}}{a}-  \frac{\dot{a}^2}{a^2}$ as 
\begin{eqnarray}
\frac{\dot{a}^2}{a^2} + \frac{k}{a^2} = \alpha \left[  \frac{\ddot{a}}{a}- \frac{\dot{a}^2}{a^2} \right] + \beta \frac{\dot{a}^2}{a^2} + \frac{1}{6}\dot{\theta}^2. \label{eqdgo2}
\end{eqnarray}
We now want to make some considerations about the field equation associated with the scalar field  given in Eq. (\ref{field2}). The FRW metric yields  $\,^{\star}\! R^{\rho\sigma\mu\nu} R_{\sigma\rho\mu\nu}=0$, thus Eq. (\ref{field2}) reads 
\begin{eqnarray}
g^{\mu \nu} \nabla_{\mu} \nabla_{\nu} \theta = g^{\mu \nu} \left[- \Gamma_{\mu \nu}^{\rho}\partial_{\rho}\theta+ \partial_{\mu} \partial _{\nu} \theta  \right] = 0. \label{nabla}
\end{eqnarray}
By choosing $\theta = \theta(t)$ \cite{3cs}, Eq. (\ref{nabla}) leads to:
\begin{eqnarray}
\ddot{\theta} + 3\frac{\dot{a}}{a}\dot{\theta} =0\,, \label{tetat}
\end{eqnarray}
what implies that \begin{eqnarray}
\dot{\theta} = C a^{-3}, \label{dotteta}
\end{eqnarray}
where $C$ represents a constant of integration. Substituting it into Eq. (\ref{eqdgo2}), we obtain the following expression:
\begin{eqnarray}
\alpha \frac{\ddot{a}}{a} + \left(\beta - \alpha -1 \right) \frac{\dot{a}^2}{a^2} - \frac{k}{a^2}  + \frac{D_1}{a^6} =0, \label{diffago}
\end{eqnarray}
where $D_1= \frac{C^2}{6}$. 
For the sake of simplicity, the following change of variable \begin{eqnarray}
u\left( a \right) = \frac{da}{dt}, \label{udia}
\end{eqnarray}
leads to  find $t$ in terms of the scale factor:
\begin{eqnarray}
t = \int u\left(a\right) ^{-1}da.  \label{intt}
\end{eqnarray}
Hence, Eq. (\ref{udia}) provides Eq. (\ref{diffago}) to be rewritten as follows:
\begin{eqnarray}
\alpha u \frac{du}{da} + \left(\beta - \alpha -1 \right)  \frac{u^2}{a} -\frac{k}{a}  + \frac{D_1}{a^5} = 0. \label{diffago2}
\end{eqnarray}
An expression of $u\left(a \right)$ is not integrable according to Eq. (\ref{intt}), and therefore we consider the case corresponding to $k=0$, a flat Universe. Thus, Eq. (\ref{diffago2}) can we written as follows:
\begin{eqnarray}
\alpha u \frac{du}{da} + \left(\beta - \alpha -1 \right)  \frac{u^2}{a} + \frac{D_1}{a^5} = 0\,,    \label{diffago3}
\end{eqnarray}
whence we can find the general solution
\begin{eqnarray}
u\left(a\right) =  a^{-2}\sqrt{  \frac{D_1+C_1\left( 1+3\alpha - \beta  \right) a^{\frac{2\left(1+3 \alpha - \beta\right)}{\alpha }}  }{1+3\alpha -\beta} },  \label{soluago}
\end{eqnarray}
where $C_1$ represents a constant of integration. 
Using Eq. (\ref{intt}), it reads
\begin{eqnarray}
t =  \sqrt{1\!+\!3\alpha\! -\!\beta} \int\!\! \frac{a^{2}}  {\sqrt{D_1\!+\!C_1\left(1\!+\!3\alpha\!-\!\beta\right) a^{\frac{2\left(1+3 \alpha - \beta\right)}{\alpha }}}}  da \label{tattabum}
\end{eqnarray}
which leads to the following solution:
\begin{eqnarray}
t&=&  \frac{a^3}{3}\sqrt{\frac{1\!+\!3\alpha \!-\!\beta}{D_1}} \nonumber\\&&\times  _2F_1\left[\frac{1}{2}, v_{\alpha,\beta}\!,1+v_{\alpha,\beta},\frac{(\beta\!-\!3\alpha \!-\!1) a^{\frac{2\left(1\!+\!3 \alpha \!-\! \beta\right)}{\alpha }} C_1}{D_1}\right], \label{tsolutiongo}
\end{eqnarray}
with $_2F_1$ representing the hypergeometric function and $v_{\alpha,\beta}\equiv \frac{3 \alpha }{2\left(1\!+\!3 \alpha-\beta\right)}$. Some considerations about the values assumed by the parameters in the hypergeometric function can be asserted, in order to find a possible analytical solution. 
We can analyze the result obtained in Eq. (\ref{tsolutiongo}) by writing the hypergeometric function as a hypergeometric series:
\begin{eqnarray}
_2F_1\left[ \frac{1}{2}, \frac{1}{2}, \frac{3}{2}, -const\times x^2 \right] = \frac{\arcsin\!{\rm h} \left(  \sqrt{const} \times x \right)}{\sqrt{const}\times x }   \label{hyper}.
\end{eqnarray}
These new results are prominent 
solutions, and the limiting case described in Eq. (\ref{hyper}) is obtained in Eq. (\ref{tsolutiongo}) when $\beta=1$, leading to the well known results in \cite{silva,myung}:
\begin{eqnarray}
 t   = \frac{1}{3\sqrt{C_1}} \arcsin\!{\rm h}\left[\sqrt{\frac{3\alpha C_1}{D_1}} a^3\left(t \right)\right].  \label{finaltgo}
\end{eqnarray}
In this way, the final expression of the scale factor 
\begin{eqnarray}
a\left(t \right)   =  \left(  \frac{C^2}{18\alpha C_1} \right)^{\frac{1}{6}}  \sinh ^{\frac{1}{3}} \left( 3\sqrt{C_1} t  \right)\,, \label{finalago}
\end{eqnarray}\noindent yields the results \cite{16cs,17cs} in this limiting case.

\subsection{The MHRDE Model}
We consider now the holographic cosmological model with IR cut-off given
by the modified Ricci radius so that the respective  energy density  is a combination of $H^2$ and $\dot{H}$ \cite{52cs,pasqua11,pasqua22}, which reads:
\begin{eqnarray}
\mathring{\rho}_{MHRDE} = \frac{2}{\alpha - \beta} \left(\dot{H} + \frac{3\alpha}{2}H^2   \right) ,   \label{MHRDE}
\end{eqnarray}
where $\alpha$ and $\beta$ are two constant parameters.
In the limiting case corresponding to ($\alpha = 4/3, \beta = 1$) we obtain that $\mathring{\rho}_{MHRDE}$ becomes proportional to
the Ricci scalar curvature $R$ for a spatially flat FRW space-time. The use of the MHRDE is motivated by the holographic principle because
it is possible to relate the DE with an UV cut-off for the vacuum energy with an IR scale such as the one
given by the Ricci scalar curvature $R$. 

By a similar procedure as in the previous Section, we obtain the following differential equation:
\begin{eqnarray}
\frac{2}{3\left( \alpha - \beta  \right)}u \frac{du}{da} + \frac{3\beta-2}{3\left( \alpha - \beta  \right)}\frac{u^2}{a} -\frac{k}{a}  + \frac{D_2}{a^5} =0, \label{umhrde}
\end{eqnarray}
where $D_2=D_1 =\frac{C^2}{6}$, which can be also written as follows:
\begin{eqnarray}
u \frac{du}{da} + A\frac{u^2}{a} -B\frac{k}{a}  + \frac{D_3}{a^5} =0, \label{umhrde2}
\end{eqnarray}
where $A=\frac{3\beta-2}{2}$, $B=\frac{3\left( \alpha - \beta  \right)}{2}$ and $D_3=\frac{C^2\left( \alpha - \beta  \right)}{4}$. 
Our calculations are severely simplified by considering a flat Universe ($k=0$), since the case with $k \neq 0$ leads to equations which cannot be solved analytically. Thus, Eq. (\ref{umhrde2}) can be rewritten as
\begin{eqnarray}
u \frac{du}{da} + A\frac{u^2}{a}  + \frac{D_3}{a^5} =0\,, \label{umhrde3}
\end{eqnarray}
which has  the following solution:
\begin{eqnarray}
u\left( a \right)  = a^{-2}\sqrt{\frac{C_2\left( A-2  \right)a^{4-2A}-D_3}{A-2}}, \label{umhrdesolution}
\end{eqnarray}
where $C_2$ is a constant of integration.

It is thus possible to find now a relation between $t$ and the scale factor:
\begin{eqnarray}
t &=&  \sqrt{A-2} \int \frac{a^2}{\sqrt{-D_3+C_2 \left( A-2  \right) a^{4-2A}}}\,da \nonumber \\
 &=& \frac{a^{1+2A}  \sqrt{ AC_2\left( A-2 \right)a^{-2A} -AD_3a^{-4}} }{\left( A+1  \right)C_2\sqrt{A\left(  A-2 \right)} }\,\,\nonumber\\&&\times  _2F_1 \left[ 1, 1+ \frac{3}{2\left( A-2 \right)}, \frac{3\left(  A-1 \right)}{2\left( A-2 \right)},  \frac{D_3a^{2A-4}}{C_2\left(  A-2 \right)}  \right]. \label{tmhrde}
\end{eqnarray}
The case corresponding to $A=5$ is going to be regarded for the sake of simplicity, as it is the single case that brings forth an analytical solution. It  implies that $\beta = 4$ and $ D_3=\frac{C^2\left( \alpha - 4  \right)}{4}$, and Eq. (\ref{tmhrde}) leads to the following expression:
\begin{eqnarray}
t = \frac{\sqrt{C_2}}{D_3} \sqrt{1-\frac{D_3a^6}{3C_2}} \,,
\end{eqnarray}
what provides the scale factor 
\begin{eqnarray}
a\left( t \right) =\left[  \frac{12C_2}{C\left( \alpha -4  \right)}  - \frac{3C\left( \alpha -4  \right)}{4}t  ^2   \right]^{\frac{1}{6}}  .\label{amhrdefinalissimo}
\end{eqnarray}
Eq. (\ref{amhrdefinalissimo}) implies that $\alpha \neq 4$ in order to avoid singularities. Moreover for  $ \frac{12C_2}{C^2\left( \alpha -4  \right)}\ll1$ we obtain $a\left(  t \right) \propto t^{\frac{1}{3}}$.

\subsection{Model with Higher Derivatives of the Hubble Parameter $H$}

We now consider a DE model proposed in \cite{34cs},  containing three different terms:  one proportional to the squared Hubble parameter, one to the first derivative with respect to the cosmic time of the Hubble parameter and another proportional to the second derivative with respect to the cosmic time of the Hubble parameter:
\begin{eqnarray}
\mathring{\rho}_{higher} =3\left( \alpha \frac{\ddot{H}}{H} + \beta \dot{H} + \gamma H^2\right),    \label{higher}
\end{eqnarray}
where $\alpha$, $\beta$ and $\gamma$ are  arbitrary dimensionless parameters.  Such  model  can be reduced to the dark
energy and Ricci-like dark energy models, for instance \cite{pasqua33}. The profile of the dark energy and the expansion of the Universe depends on the parameters $\alpha, \beta, \gamma$ of the model. The main motivation regarding this model resides on the alleviation of the age problem of three old objects, namely, LBDS
53W091,  APM 08279+5255, LBDS 53W069  \cite{age,age1,age2} for the chosen parameters. The energy density given
in Eq. (\ref{higher}) can be considered as an extension and generalization of other two DE models widely
studied recently, i.e. the Ricci DE (RDE) model and the DE energy density with Granda-Oliveros cut-off. In fact, in the limiting case corresponding to $\alpha = 0$, we obtain the energy density
of DE with Granda-Oliveros cut-off, and in the limiting case corresponding to $\alpha = 0$, $\beta = 1$ and
$\gamma = 2$, we recover the RDE model for flat Universe as well. 

The approach to this model is slightly different  to the previous two models. By substituting Eqs. (\ref{dotteta}) and (\ref{higher})  in Eq. (\ref{frinew}),  along with the new variable $x=\ln a$, the following differential equation for $H^2$ is obtained:
\begin{eqnarray}
\left(\frac{\alpha}{2}\frac{d^2}{dx^2} + \frac{\beta}{2}\frac{d}{dx} + \left(\gamma - 1   \right)\right)H^2 - ke^{-2x} +D_4 e^{-6x} =0,  \label{diffhigher}
\end{eqnarray}
where $D_4=D_1=\frac{C^2}{6}$ and $\frac{d}{dx} = H \frac{d}{dt}$.
The limiting case corresponding to $k=0$ is concerned in order to have analytical solutions of the quantities involved. Hence, Eq. (\ref{diffhigher}) can be written as follows:
\begin{eqnarray}
\frac{\alpha}{2}\frac{d^2H^2}{dx^2} + \frac{\beta}{2}\frac{dH^2}{dx} + \left(\gamma - 1   \right)H^2  +D_4 e^{-6x} =0\, \label{diffhigher2}
\end{eqnarray}
having as a solution
\begin{eqnarray}
H^2\left( x \right) &=& \frac{D_4e^{-6x}}{1-18\alpha + 3\beta - \gamma} + C_3e^{-\frac{\beta+\sqrt{\beta^2 - 8\alpha \left( \gamma-1\right) }}{2\alpha}x} \nonumber\\ &&\;\;  + C_4e^{-\frac{\beta-\sqrt{\beta^2 - 8\alpha \left( \gamma-1\right) }}{2\alpha}x}  \label{diffhigher3}
\end{eqnarray}
Passing back from $x$ to $a$, Eq. (\ref{diffhigher3}) reads\begin{eqnarray}
H^2 &=& \frac{D_4a^{-6}}{1-18\alpha + 3\beta - \gamma} + C_3a^{-\frac{\beta+\sqrt{\beta^2 - 8\alpha \left( \gamma-1\right) }}{2\alpha}} \nonumber\\&&   + C_4a^{-\frac{\beta-\sqrt{\beta^2 - 8\alpha \left( \gamma-1\right) }}{2\alpha}}. \label{diffhigher4}
\end{eqnarray}
Moreover, using now Eq. (\ref{udia}) in Eq. (\ref{diffhigher4}) yields
\begin{eqnarray}
u\left(a\right) &=& \left(\frac{D_4a^{-4}}{1-18\alpha + 3\beta - \gamma} + C_3a^{-\frac{\beta-4\alpha+\sqrt{\beta^2 - 8\alpha \left( \gamma-1\right) }}{2\alpha}}\right.\nonumber\\&&\left.    + C_4a^{-\frac{\beta-4\alpha-\sqrt{\beta^2 - 8\alpha \left( \gamma-1\right) }}{2\alpha}}\right)^{\frac{1}{2}}. \label{uhigherlim2}
\end{eqnarray}
Two different limiting cases of Eq. (\ref{uhigherlim2}) lead to two different solutions. In the first case, we choose $\beta^2 - 8\alpha \left( \gamma-1\right) =0$ and $C_3=-C_4$, what makes Eq. (\ref{uhigherlim2}) to assume the following expression:
\begin{eqnarray}
u\left( a \right) &=& \frac{1}{a^2}\sqrt { \frac{D_4}{1-18\alpha + 3\beta - \gamma}}\;\;\; \nonumber\\&&\Longrightarrow \;t = \frac{a^3}{3}\sqrt { \frac{1-18\alpha + 3\beta - \gamma}{D_4}}. \label{thigherlim1}
\end{eqnarray}
We can hence obtain the expression of the scale factor from Eq. (\ref{thigherlim1}):
\begin{eqnarray}
a\left(t\right)  = \left[ \frac{C^2}{6\left(1-18\alpha + 3\beta - \gamma \right)}\right]^{\frac{1}{6}} \left( 3t   \right)^{\frac{1}{3}} .  \label{asolhigher1}
\end{eqnarray}
In the second case, besides assuming that $\beta^2 - 8\alpha \left( \gamma-1\right) =0$,  we also consider  the case corresponding to $C_3=C_4$. Thus Eq. (\ref{uhigherlim2}) yields 
\begin{eqnarray}
u\left(a\right) =\sqrt{\frac{D_4a^{-4}}{1-18\alpha + 3\beta - \gamma} +2a^{2-\frac{\beta}{2\alpha}}}, \label{uhigherlim2new}
\end{eqnarray}
 leading to 
\begin{eqnarray}
t&=&\frac{a^3}{3} \sqrt{1\!-\!18\alpha \!+\! 3\beta \!-\! \gamma }{D_4}\nonumber\\& \times&_2F_1\!\left[\frac{1}{2},\frac{6\alpha}{12\alpha\!-\!\beta},1\!+\!\frac{6\alpha}{12\alpha\!-\!\beta},\!\frac{2\left(\gamma\!-\!3\beta\!+\!18\alpha\!-1\! \right)a^{6-\frac{\beta}{2\alpha}}}{D_4} \right]. \label{tttt}\nonumber
\end{eqnarray}
Here a similar limit can be regarded in Eq. (\ref{hyper}), which is recovered for $\beta=0$ irrespective of the value of $\alpha$, and we obtain the following solution:
\begin{eqnarray}
t=\frac{1}{3\sqrt{2}}\arcsin\!{\rm h} \left[ \sqrt{\frac{2\left(  1-18\alpha  - \gamma \right)}{D_4}}a^3 \right],   \label{tsolhigher2}
\end{eqnarray}
which leads to the following solution for the scale factor:
\begin{eqnarray}
a\left( t   \right) = \left[\frac{C^2}{12\left(  1-18\alpha  - \gamma \right)}\right]^{\frac{1}{6}} \sinh^{\frac{1}{3}}\left( 3\sqrt{2} t \right) . \label{asolhigher2}
\end{eqnarray}
Some considerations about the values of $\alpha$ and $\gamma$ in Eq. (\ref{asolhigher2}) can be summoned now. The condition $\beta^2 = 8\alpha\left( \gamma-1\right) $ obviously reads $8\alpha\left( \gamma-1\right) =0 $ for $\beta =0$,  implying a tricotomy: 1) $\alpha =0 $  and $\gamma \neq 1$, 2) $\alpha \neq 0 $  and $\gamma = 1$ and 3) $\alpha =0 $  and $\gamma = 1$. We analyze the respective solutions for these three conditions. 
For the case 1), 
\begin{eqnarray}
a\left( t   \right)   =  \left[\frac{C^2}{12\left(  1  - \gamma \right)}\right]^{\frac{1}{6}} \sinh^{\frac{1}{3}}  \left( 3\sqrt{2} t \right) . \label{asolhigher2-1}
\end{eqnarray}
Instead, for the case 2) we obtain the scale factor in the form 
\begin{eqnarray}
a\left( t   \right)    =  \left[\frac{C^2}{-216\alpha}\right]^{\frac{1}{6}}  \sinh^{\frac{1}{3}}  \left( 3\sqrt{2} t \right)   .  \label{asolhigher2-2}
\end{eqnarray}
Finally, for case 3), Eq. (\ref{asolhigher2}) diverges since  $1-18\alpha -\gamma =0$. This result is in agreement with the one when  $\alpha = \beta =0$ and $\gamma=1$ in Eq. (\ref{frinew}): with this combination of the parameters, we obtain the equation $Ca^{-6} =0$, which solutions are $C=0$ or $a \rightarrow   \infty$ (which is the result with the  case 3)). Thus, the combination of values considered in 3) cannot be regarded, since it does not produce an analytical  solution.

\section{Conclusions}
In this work we studied the behavior of three different DE models: the HDE model with Granda-Oliveros cut-off, the MHRDE model and the model with higher derivatives of the Hubble parameter $H$, in the framework of the Chern-Simons modified gravity model. For each of these models, we derived the respective scale factors $a\left( t \right)$.

For the HDE model with GO cut-off, the scale factor $a\left( t \right) $ is an hyperbolic sine function of cosmic time.
 Nevertheless, in the MHRDE model paradigm the scale factor is a power law of the time, and finally, according to the values of the parameters involved for the model with higher derivatives of the Hubble parameter, we have either a power law solution or $a\left( t \right) $  proportional to a hyperbolic sine function.
 
The  scale factor obtained in Eq. (\ref{finalago}) for the HDE with GO cut-off and in Eqs. (\ref{asolhigher2-1}) and (\ref{asolhigher2-2}) for the model with higher derivatives of $H$ are similar to those obtained in \cite{16cs, 17cs, silva}.
For this reason, we conclude that, for suitable choices of the parameters involved, the HDE model with GO cut-off and the model with higher derivatives of the Hubble parameter $H$ 
in the framework of Chern-Simons modified gravity have the same results obtained from the
modified Chaplygin gas \cite{cgas3,cgas41,cgas42}, namely, the results    clearly indicate that there is a agreement  between both the
the HDE model with GO cut-off and the model with higher derivatives of the Hubble parameter $H$  in the framework gravity Chern-Simons,  and the modified Chaplygin gas. It is worth  to emphasize that as Ricci dark energy in Chern-Simons modified gravity is related to Ricci dark energy with a minimally coupled scalar when choosing the FRW metric, the above mentioned similarity between them is limited to the de Sitter phase derived by the cosmological constant in the future \cite{myung}.

\begin{acknowledgements}
Financial support under Grant No. SR/FTP/PS-167/2011 from DST, Govt of India is thankfully acknowledged by the second author.
RdR is grateful to thanks to SISSA for the hospitality, to CNPq grants No. 473326/2013-2 and No. 303027/2012-6. RdR is also \emph{Bolsista da CAPES Proc. 10942/13-0.}
\end{acknowledgements}


\begin{thebibliography}{99}
\bibitem{1-1} P.~de Bernardis {\it et al.}  [Boomerang Collaboration],
  Nature, 955 {\bf 404} (2000)
\bibitem{1a} S.~Perlmutter {\it et al.}  [Supernova Cosmology Project Collaboration],
  Astrophys.\ J.\  {\bf 517}, 565 (1999).
\bibitem{1b} A.~G.~Riess {\it et al.}  [Supernova Search Team Collaboration],
  Astron.\ J.\  {\bf 116}, 1009 (1998).

\bibitem{cmb1} C. L. Bennett et al., Astrophys. J. \textbf{148}, 1 (2003).
\bibitem{planck} Planck Collaboration, P. A. R. Ade,  N. Aghanim et al.\ 2013, arXiv:1303.5076
\bibitem{tegmark}  M.~Tegmark {\it et al.}  [SDSS Collaboration],
  Astrophys.\ J.\  {\bf 606}, 702 (2004).
\bibitem{xray} S. W. Allen,  et al., Mon. Not. Roy. Astron. Soc. \textbf{353}, 457 (2004).

\bibitem{Peebles} P.~J.~E.~Peebles and B.~Ratra,
  Rev.\ Mod.\ Phys.\  {\bf 75}, 559 (2003). 
  
\bibitem{sahni1}V.~Sahni,
  Class.\ Quant.\ Grav.\  {\bf 19}, 3435 (2002).
\bibitem{copeland-2006} E.~J.~Copeland, M.~Sami and S.~Tsujikawa,
  Int.\ J.\ Mod.\ Phys.\ D {\bf 15}, 1753 (2006).
\bibitem{clift} T.~Clifton, P.~G.~Ferreira, A.~Padilla and C.~Skordis,
  Phys.\ Rept.\  {\bf 513}, 1 (2012). 


\bibitem{bambaDE} K.~Bamba, S.~Capozziello, S.~'i.~Nojiri and S.~D.~Odintsov,
  Astrophys.\ Space Sci.\  {\bf 342}, 155 (2012).


\bibitem{delcampo} S. del Campo, R. Herrera,  D. J. Pavon, \emph{JCAP} \textbf{0901}, 020 (2009).

\bibitem{delcampoh} M. Jamil, E. Saridakis, \emph{JCAP} \textbf{07}, 028 (2010).

  
 
   \bibitem{Horava} P. Ho${\check{\rm r}}$ava and D. Minic,  
  Phys.\ Rev.\ Lett.  {\bf 85}, 1610 (2000).

 \bibitem{3b} Y.~S.~Myung and M.~-G.~Seo,
  Phys.\ Lett.\ B {\bf 671}, 435 (2009).
  
\bibitem{4} Q.~-G.~Huang and M.~Li,
  JCAP {\bf 0408}, 013 (2004).
  
  
  
  
\bibitem{rami1} M.~Li, X.~-D.~Li, Y.~-Z.~Ma, X.~Zhang and Z.~Zhang,
  JCAP {\bf 1309}, 021 (2013).
  
  
\bibitem{rami3} I.~Duran and L.~Parisi,
  Phys.\ Rev.\ D {\bf 85}, 123538 (2012).
  


\bibitem{cardenas}
  V.~Cardenas and R.~G.~Perez,
  Class.\ Quant.\ Grav.\  {\bf 27}, 253003 (2010).
  
\bibitem{wu2008}
  S.~-F.~Wu, P.~-M.~Zhang and G.~-H.~Yang,
  Class.\ Quant.\ Grav.\  {\bf 26}, 055020 (2009).
  
\bibitem{zim}
  W.~Zimdahl and D.~Pavon,
  Class.\ Quant.\ Grav.\  {\bf 24}, 5461 (2007).


\bibitem{0505}
  Y.~-g.~Gong and Y.~-Z.~Zhang,
  Class.\ Quant.\ Grav.\  {\bf 22}, 4895 (2005).
  

\bibitem{3a} Y.~S.~Myung,
  Phys.\ Lett.\ B {\bf 649}, 247 (2007).
  
  
  
\bibitem{5a} L.~Susskind,
  J.\ Math.\ Phys.\  {\bf 36}, 6377 (1995).
  
  \bibitem{li997} M.~Li, X.~-D.~Li, S.~Wang, Y.~Wang and X.~Zhang,
  JCAP {\bf 0912}, 014 (2009).
  
\bibitem{li998} M.~Li, X.~-D.~Li, S.~Wang and X.~Zhang,
  JCAP {\bf 0906}, 036 (2009).




  
\bibitem{gong2} Q.~-G.~Huang and Y.~-G.~Gong,
  JCAP {\bf 0408}, 006 (2004).
  
\bibitem{noji1} E.~Elizalde, S.~'i.~Nojiri, S.~D.~Odintsov and P.~Wang,
  Phys.\ Rev.\ D {\bf 71}, 103504 (2005).

\bibitem{noji2} X.~Zhang and F.~-Q.~Wu,
  Phys.\ Rev.\ D {\bf 72}, 043524 (2005).

\bibitem{noji3}  B.~Guberina, R.~Horvat and H.~Stefancic,
  JCAP {\bf 0505}, 001 (2005).

\bibitem{noji4}  B.~Wang, Y.~-g.~Gong and E.~Abdalla,
  Phys.\ Lett.\ B {\bf 624}, 141 (2005).

\bibitem{noji5} H.~Li, Z.~-K.~Guo and Y.~-Z.~Zhang,
  Int.\ J.\ Mod.\ Phys.\ D {\bf 15}, 869 (2006).

\bibitem{noji6} J.~P.~Beltran Almeida and J.~G.~Pereira,
  Phys.\ Lett.\ B {\bf 636}, 75 (2006).

  
  \bibitem{10}B.  Chen, M. Li, Y. Wang, Nucl.  Phys. B \textbf{774}, 256 (2007).



\bibitem{12a} H. M. Sadjadi, M. Jamil, Gen.  Relat. Grav. \textbf{43}, 1759 (2011).

\bibitem{12e} M. Jamil, M. U.  Farooq, M. A. Rashid, European Phys. J. C \textbf{61}, 471 (2009).

  
\bibitem{13a} B. Wang, C. Y. Lin,  E. Abdalla,   Phys. Lett. B \textbf{637}, 357 (2006).

\bibitem{13c} A. Sheykhi,  Class. Quant. Grav.  \textbf{27}, 025007 (2010).
 
 
  
  Astrophys. J. Suppl. {\bf 192}, 18 (2011).

  



\bibitem{14} S. Chattopadhyay, U. Debnath,  Astrophys. Space Sci., \textbf{319}, 183 (2009).

\bibitem{16} C. Feng, B. Wang, Y. Gong, R. K. Su, \emph{JCAP} \textbf{9}, 5 (2007).

\bibitem{16d} J. Lu, E. N. Saridakis, M. R.  Setare, L. Xu, \emph{JCAP} \textbf{3}, 31 (2010).


\bibitem{3cs} R. Jackiw, S. Y. Pi, Phys. Rev. D \textbf{68}, 104012 (2003).


\bibitem{taveras} V. Taveras and N. Yunes, Phys. Rev. D {\bf 78}, 064070
(2008).

\bibitem{calcagni} G. Calcagni and S. Mercuri, Phys. Rev. D {\bf 79}, 084004
(2009).


\bibitem{mercuri} S. Mercuri and V. Taveras, Phys. Rev.  D {\bf 80}, 104007
(2009).

\bibitem{rocha} R.~da Rocha and J.~G.~Pereira,
  Int.\ J.\ Mod.\ Phys.\ D {\bf 16} (2007) 1653
  

\bibitem{6cs} S. Alexander, N.  Yunes, Phys. Rep. \textbf{480}, 155 (2009).

\bibitem{13cs} D. Grumiller, N. Yunes, Phys. Rev. D. \textbf{77}, 044015 (2008).

\bibitem{csca}P.~Canizares, J.~R.~Gair and C.~F.~Sopuerta,
  Phys.\ Rev.\ D {\bf 86}, 044010 (2012).


\bibitem{7cs} K. Yagi, L. C. Stein, N. Yunes, T.  Tanaka, Phys. Rev. D \textbf{87}, 084058 (2013).



 
  \bibitem{34cs} S. Chen, J. Jing, Phys. Lett. B \textbf{679}, 144 (2009).



 \bibitem{cga} 
  M.~C.~Bento, O.~Bertolami and A.~A.~Sen,
  Phys.\ Rev.\ D {\bf 66}, 043507 (2002).

 \bibitem{cgas0} A.~Y.~.Kamenshchik, U.~Moschella and V.~Pasquier,
  Phys.\ Lett.\ B {\bf 511}, 265 (2001).

\bibitem{cgask} 
  V.~Gorini, A.~Kamenshchik and U.~Moschella,
  Phys.\ Rev.\ D {\bf 67}, 063509 (2003).
\bibitem{cgas3} M. R. Setare, European Physical Journal C \textbf{52}, 689 (2007).

\bibitem{cgas41} A. E. Bernardini, O.  Bertolami, O. Phys. Rev. D {\bf 77}, 083506 (2008).

\bibitem{cgas42} A. E. Bernardini, Astropart. Phys. {\bf 34}, 431
(2011).

 \bibitem{grandaoliveroscs}  L.~N.~Granda and A.~Oliveros,
  Phys.\ Lett.\ B {\bf 671}, 199 (2009).
    
\bibitem{grandaoliverosacs} L.~N.~Granda and A.~Oliveros,
  Phys.\ Lett.\ B {\bf 669}, 275 (2008).


\bibitem{silva} J. G. Silva, A. F. Santos,  European Phys. J. C \textbf{73}, 2500 (2013). 

\bibitem{myung} 
  Y.~S.~Myung,
  Eur.\ Phys.\ J.\ C {\bf 73}, 2515 (2013).

\bibitem{16cs} S. S. Costa, Mod. Phys. Lett. A \textbf{24}, 531 (2009).

\bibitem{17cs} J. C. Fabris, S. V. B. Goncalves, R. de Sa Ribeiro, Gen. Rel. Grav. \textbf{36}, 211 (2004).

  
  










\bibitem{52cs} L. P. Chimento, M. G. Richarte,  Phys. Rev. D \textbf{85}, 127301 (2012).
\bibitem{pasqua11} A. Jawad, S. Chattopadhyay, A.    Pasqua,  Eur. Phys. J. Plus, \textbf{128}, 88 (2013).

\bibitem{pasqua22}
  A.~Pasqua, S.~Chattopadhyay and I.~Khomenko,
  Can.\ J.\ Phys.\  {\bf 91}, 632 (2013). 

\bibitem{pasqua33} 
  A.~Pasqua and S.~Chattopadhyay,
  Int.\ J.\ Theor.\ Phys.\  {\bf 53}, 435 (2014).

  
  \bibitem{age}
 J. Dunlop, et al., Nature {\bf 381}, 581 (1996).
\bibitem{age1} H. Spinrad, et al., Astrophys. J. {\bf 484}, 581 (1999).
\bibitem{age2} D. Jain, A. Dev, Phys. Lett. B {\bf 633},  436 (2006).
  







  







 
  







\end{thebibliography}
\end{document}